\documentclass[12pt,preprint]{aastex}
\shorttitle{Optimal Chandra/XMM bands}
\shortauthors{Scharf}

\begin{document}
\title{Optimal Chandra/XMM-Newton band-passes for detecting low 
temperature groups and 
clusters of galaxies}

\author{Caleb Scharf}
\affil{Columbia Astrophysics Laboratory, Columbia University, MC5247, 550 
West 120th St., New York, NY 10027, USA.}
\email{caleb@astro.columbia.edu}

\begin{abstract} In this short paper I present the results of a
calculation which seeks the maximum, or optimal, signal-to-noise energy
band for galaxy group or cluster X-ray emission detected by the Chandra
and XMM-Newton observatories. Using a background spectrum derived from
observations and a grid of models I show that the "classical" 0.5-2 keV
band is indeed close to optimal for clusters with gas temperatures $>2
$ keV, and redshifts $z<1$. For cooler systems however, this band is
generally far from optimal.  Sub-keV plasmas can suffer 20-60\%
signal-to-noise loss compared to an optimal band, and worse for $z>0$. The
implication is that current and forthcoming surveys should be carefully
constructed in order to minimize bias against the low mass, low
temperature end of the cluster/group population.

\end{abstract}


\section{Introduction} The X-ray emission from $10^6-10^8$K plasma in the
gravitational potentials of massive clusters and groups of galaxies has
proved to be an invaluable means by which these systems can be robustly
detected and quantified, e.g. \citet{gio90}. The use of all- or
partial-sky X-ray data \citep{ebe97,hen92,gio90} and archival pointed data
\citep{ros95,sch97,vik98} in constructing statistically complete catalogs
of clusters and groups has dramatically filled in our picture of their
evolution, from low redshifts to $z\sim 1$ \citep{bor01}. The new Chandra
and XMM-Newton observatories promise to further extend this early work,
and in combination with measurements of the Sunyaev Z'eldovich (S-Z) CMB
decrement, will allow us to refine and extend our use of clusters as tools
for cosmology \citep{hai01}. At the lower end of the mass scale, Chandra
and XMM-Newton will permit the poorer, lower luminosity clusters and
groups (systems with temperatures $<2$ keV) to be systematically detected
for the first time to redshifts of a few tenths. Previous surveys for
these systems have often had to rely on prior optical catalogs,
complicating the estimation of statistical completeness and bias
\citep{mul98}.

A common aspect of X-ray group and cluster surveys has been the frequent
(although not universal) choice of a pass-band from 0.5-2keV. Such a band
typically helps minimize the contribution of soft Galactic emission, and
harder particle background, while maximizing sensitivity to general
cluster emission \citep{mch98,ros95}. Previous imaging X-ray
observatories have also not had the good spectral resolution of Chandra or
XMM-Newton, so the choice of band was less controllable.

In this short paper I present a set of simple calculations aimed at
determining a set of optimal pass-bands for thermal plasmas over a range
of temperatures and redshifts. I also demonstrate the relationship to the
commonly used 0.5-2 keV band, and argue that future surveys of cooler
systems must take the band-pass explicitly into account, or risk seriously
biasing their estimates of the space density of such systems.

\section{Optimal energy bands}

The signal-to-noise criterion I use here is $S/\sqrt N$, where $S$ is the
source photon count, $N$ is the appropriate background count. This is
chosen as a decent approximation to the various forms of detection
significance criteria used in X-ray surveys for extended sources.
Typically, a mean background count per sky area is estimated, and the
deviation of the count rate in a given area, minus the background, is
compared to the statistical (usually Poisson) fluctuation in the
background over that area.

\subsection{The calculations}

A typical background spectrum is derived across the full instrument
band-pass by combining high Galactic latitude data ($<nH>\sim 1\times
10^{21}$) in which bright sources and periods of high particle background
(flares) have been removed. For Chandra, the data was culled by combining
archival fields and the online background data. Similarly, for XMM,
archival PV data and online background data was utilized. The background
spectrum is derived by binning photons from the entire field-of-view and
performing a simple, linear, sliding window smoothing over 5 PI channels.
The spectra so derived are still somewhat noisy, but not at a level which
alters the results presented here.

A grid of models is generated across a range of temperatures and redshifts
using XSPEC, with a MEKAL plasma, 1/3rd Solar abundances, $<nH>$ is set to a
high Galactic latitude maximum of $1\times 10^{21}$cm$^2$, and the results are
robust to reasonable variations in these quantities. $\Delta z$ is set to $0.1$
and $kT$ is actually set over a range from 0.1 to 12.9 keV in steps of $0.1$
keV, although only temperatures of 0.2, 0.5, 1.0, 2.0, 4.0, and 6.0 keV are
presented here.  The spectra are then forward-folded through the on-axis RMF's
and ARF's for the respective instruments. In the case of the Chandra
front-illuminated CCD device (ACIS-I), off-axis responses are also used in an
effort to evaluate the effect of radiation damage induced charge-transfer
inefficiencies (CTI) (see \S 2.2 \&3). All Chandra responses used here assume
that basic CTI corrections have beeen applied to the data (e.g. \citet{tow00}).
The corrections reduce the position dependence (on the chips) of the gain and
grade distributions (which distort the inferred photon energies), but an energy
and position dependent degradation of spectral resolution remains.

An unrestricted search is then made for the maximal signal-to-noise by
varying the lower ($E_1$) and upper ($E_2$) energy bands independently,
as a function of $z$.

\subsection{The results}

Figures 1,2,3, and 4 present the optimal band search results for, 
respectively, the Chandra ACIS-I, ACIS-S, the XMM-Newton MOS, and PN
instruments.

Some of the energy band limits show clear features as a function of
redshift, however these all correspond to small variations in actual
signal-to-noise, as is reflected by the smoothness of the 0.5-2 keV to
optimal ratio curves. The features are due to the combination of the shape
of the instrumental response and features in the background and source
spectra. For example, in Figure 2, the upper limit band pass curve for
$kT=6$ keV (heaviest line) exhibits a sharp drop as the spectrum redshifts
from $z=0.6$ to 0.7. This is entirely due to the location of a flourescent
Si K-$\alpha$ line (at $\sim 1.7$ keV) in the background, from reflection
in the Chandra mirror assembly. As the redshift of the source increases a
critical point is reached where the optimal band edge crosses this
background line, and the optimum jumps to a lower energy.

The key results may be summarized as follows. For all instruments, plasmas with
$kT<2$ keV have optimal bands which differ significantly from the 0.5-2 keV
bandpass. For the coolest plasma considered here ($0.2$ keV, $2.3\times 10^6$
K) at $z=0.1$ the 0.5-2 keV band suffers a 55-65\% S/N loss with Chandra, and a
$\sim 45$\% S/N loss with XMM compared to the optimal band pass. The optimal
band in this case has a maximum range from $0.4-0.7$ keV for Chandra and
$0.4-0.8$ for XMM, both of which are very narrow. At higher redshifts the S/N
loss for the 0.5-2 keV band increases significantly. As the plasma temperature
increases the optimal band rapidly widens. For a 1 keV plasma at $z=0.1$ the
optimal bands are in the range of $0.6-1.2$ keV, and have a S/N only 10\%
better than 0.5-2 keV, although this typically increases with redshift. The
results for a 6' off-axis response function for the Chandra ACIS-I instrument,
with a lower CCD row number and hence smaller CTI, are very similar to the high
row number, on-axis calculations. Variations in band limits between the two are
$\leq 5$\%, and therefore negligible.

In a survey, where the space density is to be recovered, the effective volume
in which a source of a given intrinsic luminosity can be detected, is
calculated based on the maximal redshift of detectability. A 10\% drop in S/N
compared to that expected propagates into a $\sim 10$\% error in volume,
consequently the band corrections suggested here are critical.

\section{Conclusion}

With luminosities of $10^{41-43}$erg s$^{-1}$, low mass, cool ($kT<2$
keV) groups and clusters of galaxies will form a significant fraction of
the extended emission X-ray systems detectable to $z\sim 0.5$ in
medium-deep Chandra and XMM exposures. Probing this population of
collapsed systems is vital for improving our understanding of both the
overall cluster mass function and the regime where gravitational collapse
and astrophysical energies are comparable \citep{llo00}.

I have demonstrated here that the choice of band-pass is critical in both
maximizing the detection sensitivity for clusters, and in its correct
quantification in order to recover the true space density of poor cluster
systems. The optimal bands presented here can serve as a reference point for
Chandra and XMM surveys.

For the radiation damaged ACIS-I on Chandra, once the data is processed to
correct for much of the CTI effect (e.g. \citet{tow00}) the impact of the
reduced spectral resolution on the optimal band limits is minimal (similarly
for the inherent CTI in the back illuminated chips), and $\leq 5$\% in
amplitude at all energies. It appears that it can therefore be safely ignored
in this situation.

It has been assumed here that cluster emission is isothermal. In reality
this is often not the case, and significant temperature structure or
gradients are present in massive clusters, and are likely also in lower
mass systems. This will further complicate not only the absolute detection
of a cluster, but could bias the estimation of flux in a single band.  In
a future work (Scharf, in preparation) I will discuss the more complex
issues involved with X-ray and S-Z detection biases for clusters in the
context of their use as cosmological probes.

\acknowledgements CAS gratefully acknowledges helpful discussions with D.  
Helfand and F. Paerels and the generous support of the Columbia
Astrophysics Laboratory for this work.

\begin{figure}
\caption{Results of an optimal band search for the
Chandra ACIS-I FI chip array. Left panel (a): Upper (solid) and lower
(dashed) energy band limits ($E_2$ and $E_1$) are plotted as a function of
redshift ($z$).  Plasma temperatures are 0.2 keV (lightest weight curves),
0.5 keV, 1 keV, 2 keV, 4 keV, and 6 keV (heaviest curves). Right panel
(c): The ratio of signal-to-noise in the classical 0.5-2 keV band to that
in the optimal band is plotted (as a percentage) versus redshift for the 6
plasma temperatures.}
\plotone{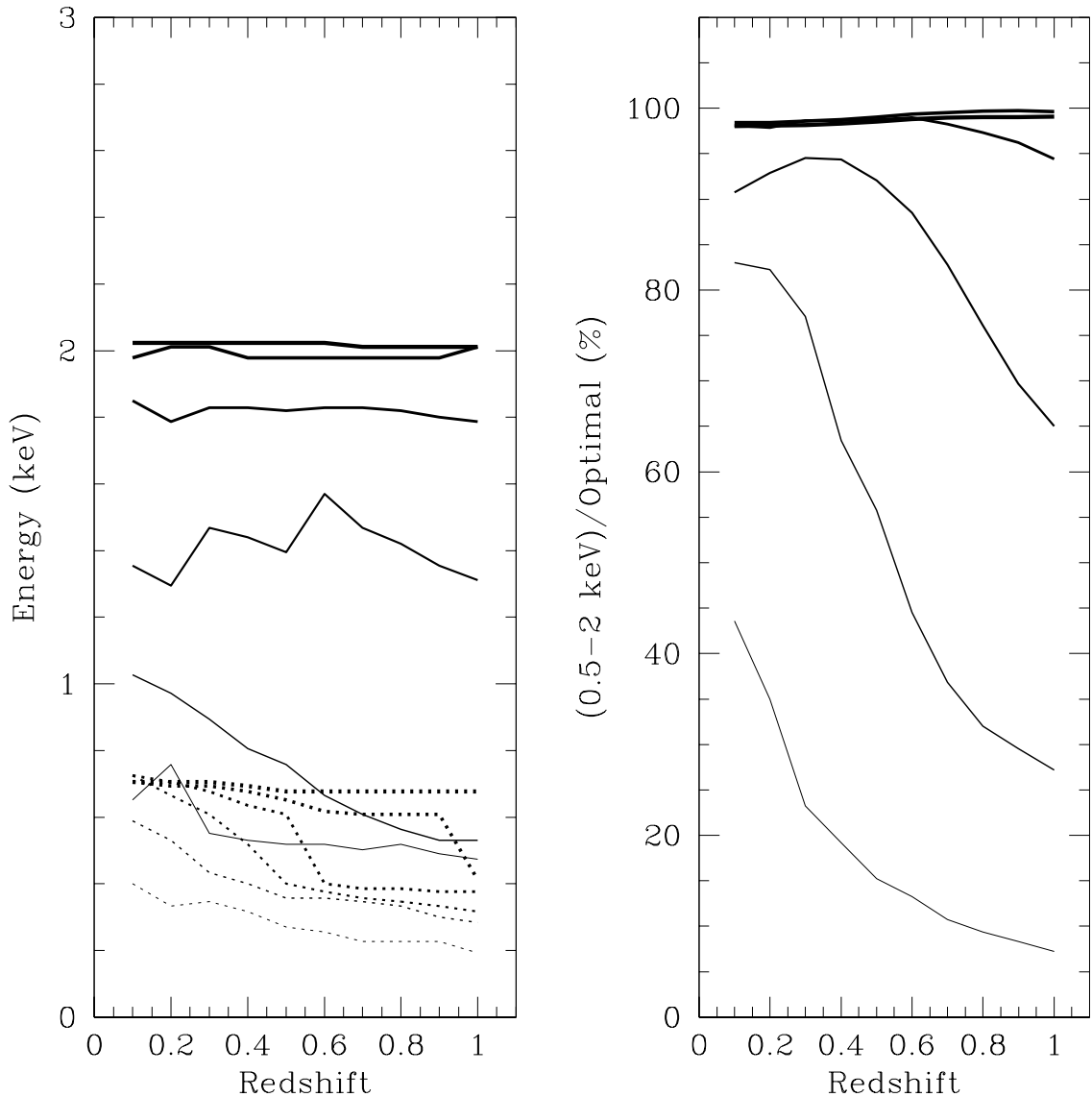}
\end{figure}

\begin{figure}
\caption{As for Figure 1, for the Chandra ACIS-S BI chips}
\plotone{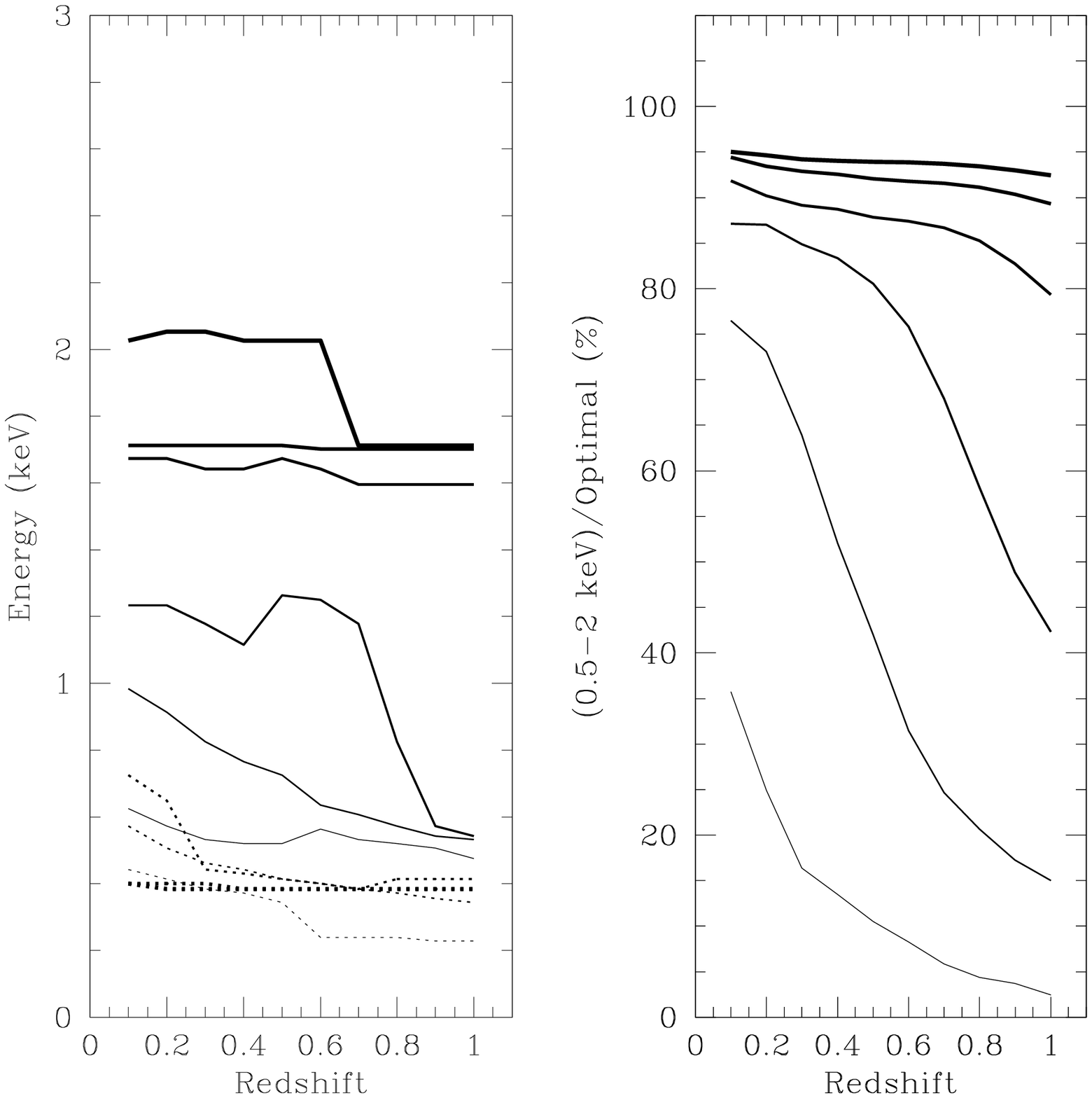}
\end{figure}

\begin{figure}
\caption{As for Figures 1 \& 2, but for the XMM-Newton MOS 
instrument with a thin optical blocking filter}
\plotone{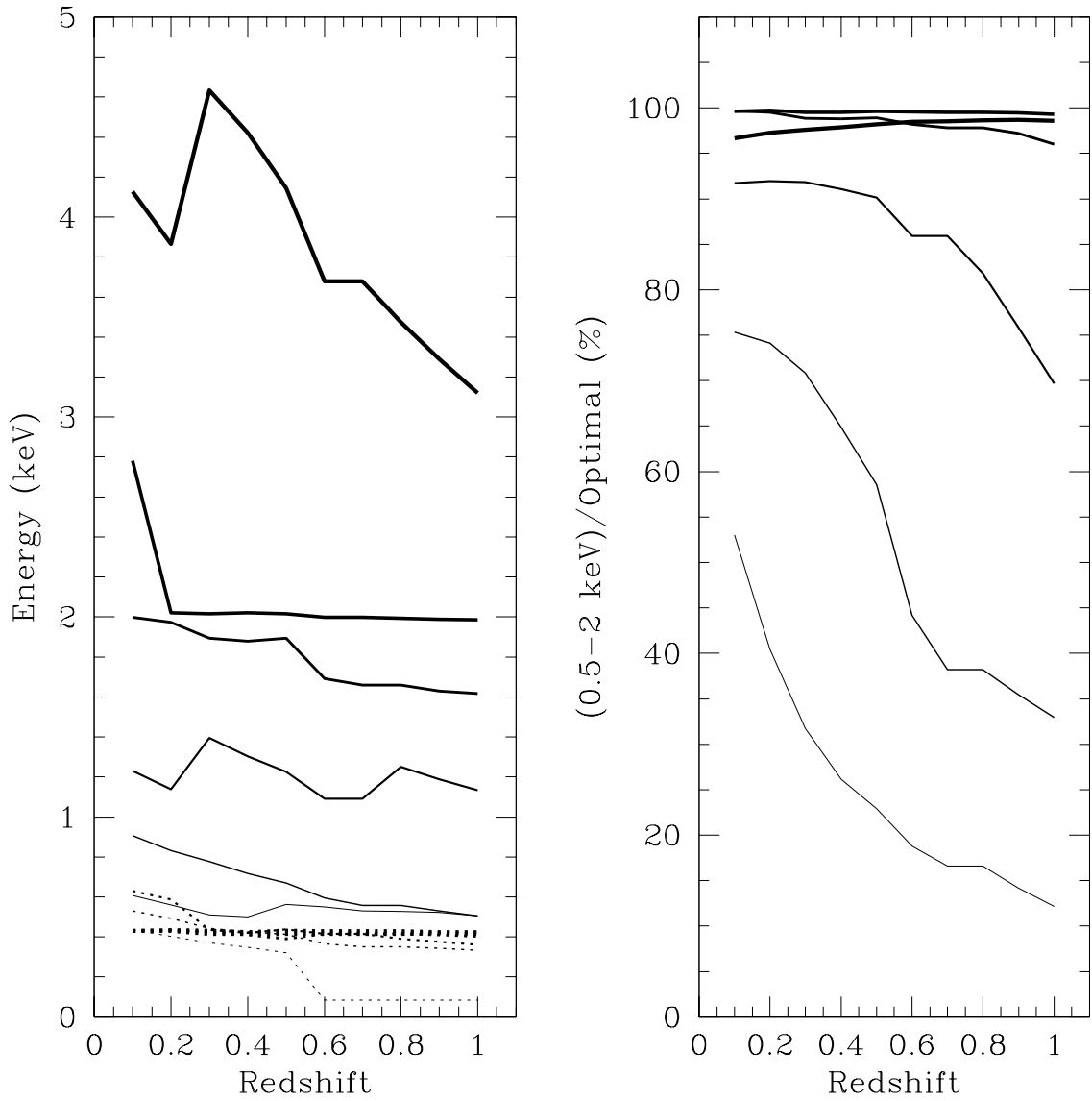}
\end{figure}

\begin{figure}
\caption{As for Figure 3, for the XMM-Newton PN instrument 
with a thin optical blocking filter}
\plotone{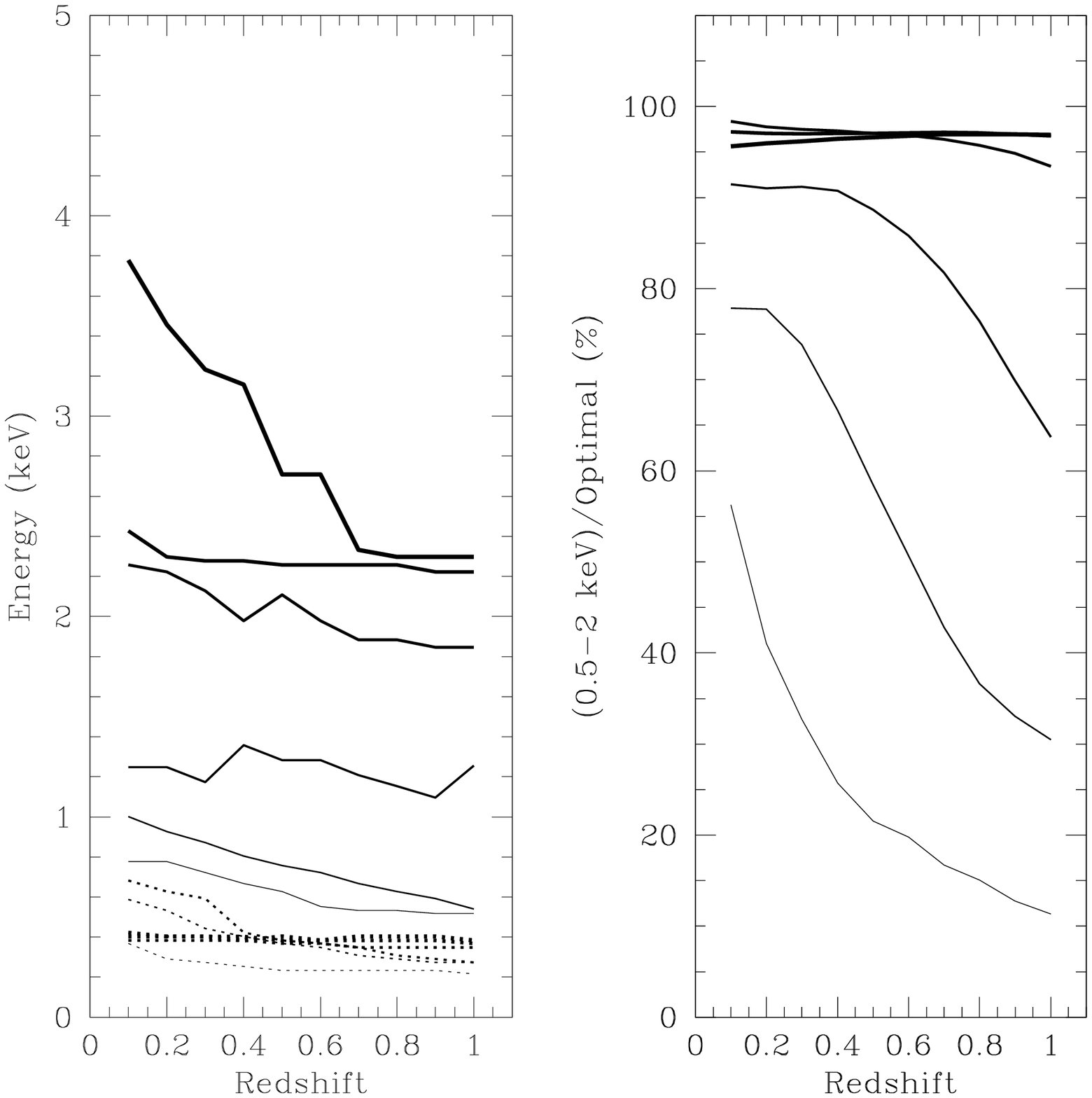}
\end{figure}

\end{document}